\documentclass[%
 reprint,
 amsmath,amssymb,
 aps,
 prl,
]{revtex4-2}

\usepackage{graphicx}
\usepackage{dcolumn}
\usepackage{bm}
\usepackage{upgreek} 
\usepackage[colorlinks,linkcolor=blue,urlcolor=blue,citecolor=blue]{hyperref}

\begin{document}

\preprint{APS/123-QED}

\title{Attosecond Compression of Relativistic Electron Pulses via Continuous Harmonic Undulator Resonance}

\author{Hao Sun\textsuperscript{1}}
\author{Xiaofan Wang\textsuperscript{1}}
\email{wangxf@mail.iasf.ac.cn}
\author{Xiaozhe Shen\textsuperscript{1}}
\email{shenxiaozhe@mail.iasf.ac.cn}
\author{Huaiqian Yi\textsuperscript{1}}
\author{Cheng-Ying Tsai\textsuperscript{2}}
\author{Weiqing Zhang\textsuperscript{3}}
\email{weiqingzhang@dicp.ac.cn}
\author{Xueming Yang\textsuperscript{1,3}}

\affiliation{
1. Institute of Advanced Light Source Facilities, Shenzhen, Shenzhen 518107, China\\
2. School of Electrical and Electronic Engineering, Huazhong University of Science and Technology, Wuhan, 430073, China\\
3. State Key Laboratory of Chemical Reaction Dynamics, Dalian
Institute of Chemical Physics, Chinese Academy of Sciences, Dalian 116023, China
}

\date{\today}

\begin{abstract}
Extending megaelectronvolt ultrafast electron diffraction (MeV UED) into the attosecond regime is essential for resolving intrinsic structural dynamics, yet requires simultaneously controlling electron-pulse duration and arrival-time stability. Here, we propose a generalized harmonic laser-electron interaction that extends beam modulation into a continuous harmonic regime. We demonstrate that highly detuned, non-integer harmonic modulation via a single-period undulator achieves stronger coupling efficiency than conventional integer-harmonic resonance. Driven by a mid-infrared seed laser whose wavelength is a small fraction of the nominal resonant wavelength, this mechanism enables effective longitudinal phase space manipulation. It facilitates attosecond compression with minimal laser-induced energy spread, preserving the beam quality required for high-fidelity diffraction. Furthermore, deriving both the modulation and experimental pump lasers from a common source intrinsically locks their relative timing. Simulations demonstrate 680-as pulse durations and 470-as arrival-time jitter, establishing a viable route to attosecond MeV UED for resolving coupled electron-nuclear dynamics.
\end{abstract}

             
\maketitle

Attosecond science \cite{PierreAgostini_2004,RevModPhys.81.163} has enabled direct observation of electronic dynamics on their natural timescales using attosecond light sources based on high-harmonic generation \cite{doi:10.1126/science.aao7043,jordanAttosecondSpectroscopyLiquid2020}, and free-electron lasers \cite{liAttosecondCoherentElectron2022, driverAttosecondDelaysXray2024}. A complete picture of ultrafast phenomena further requires access to the accompanying atomic structural response, which calls for probes with direct sensitivity to atomic positions. Electron diffraction enables direct structural measurements, and keV electron beams have routinely demonstrated Ångström-scale spatial resolution with sub-picosecond temporal resolution \cite{doi:10.1126/science.1248488,doi:10.1126/science.1166135}. Recent advances have further enabled attosecond control of keV electron beams, demonstrating the generation of attosecond pulse trains \cite{morimotoAttosecondElectronbeamTechnology2023} and its application to sub-optical-cycle imaging of optoelectronic dynamics through field-sensitive interaction mechanisms \cite{morimotoDiffractionMicroscopyAttosecond2018,nabbenAttosecondElectronMicroscopy2023,gaidaAttosecondElectronMicroscopy2024a}. However, strong Coulomb repulsion among electrons presents a significant challenge to simultaneously achieving such high temporal resolution and the electron flux required for high-fidelity structural imaging, particularly in complex or beam-sensitive systems. Addressing this bottleneck motivates the extension of attosecond control to the relativistic MeV electron beam regime.

At MeV energies, space-charge effects are strongly suppressed, naturally facilitating the generation and preservation of ultrashort and high-coherence electron pulses. Building upon the established success of MeV-UED in achieving Ångström-scale spatial resolution with 100-fs temporal resolution in solids \cite{moHeterogeneousHomogeneousMelting2018a,sieUltrafastSymmetrySwitch2019,soodUniversalPhaseDynamics2021,zongSpinmediatedShearOscillators2023,cheng_ultrafast_2024}, gases \cite{yangImagingCF3IConical2018, yangSimultaneousObservationNuclear2020, champenoisConformerspecificPhotochemistryImaged2021}, and liquid-phase systems \cite{yangDirectObservationUltrafast2021, linImagingShortlivedHydroxylhydronium2021}, advancing MeV-UED into the attosecond regime represents a critical step toward directly resolving coupled electron–nuclear dynamics in real time across a wide range of systems and disciplines \cite{doi:10.1021/jacs.2c07997}.

\begin{figure*}[htbp] 
\centering
\includegraphics[width=18cm]{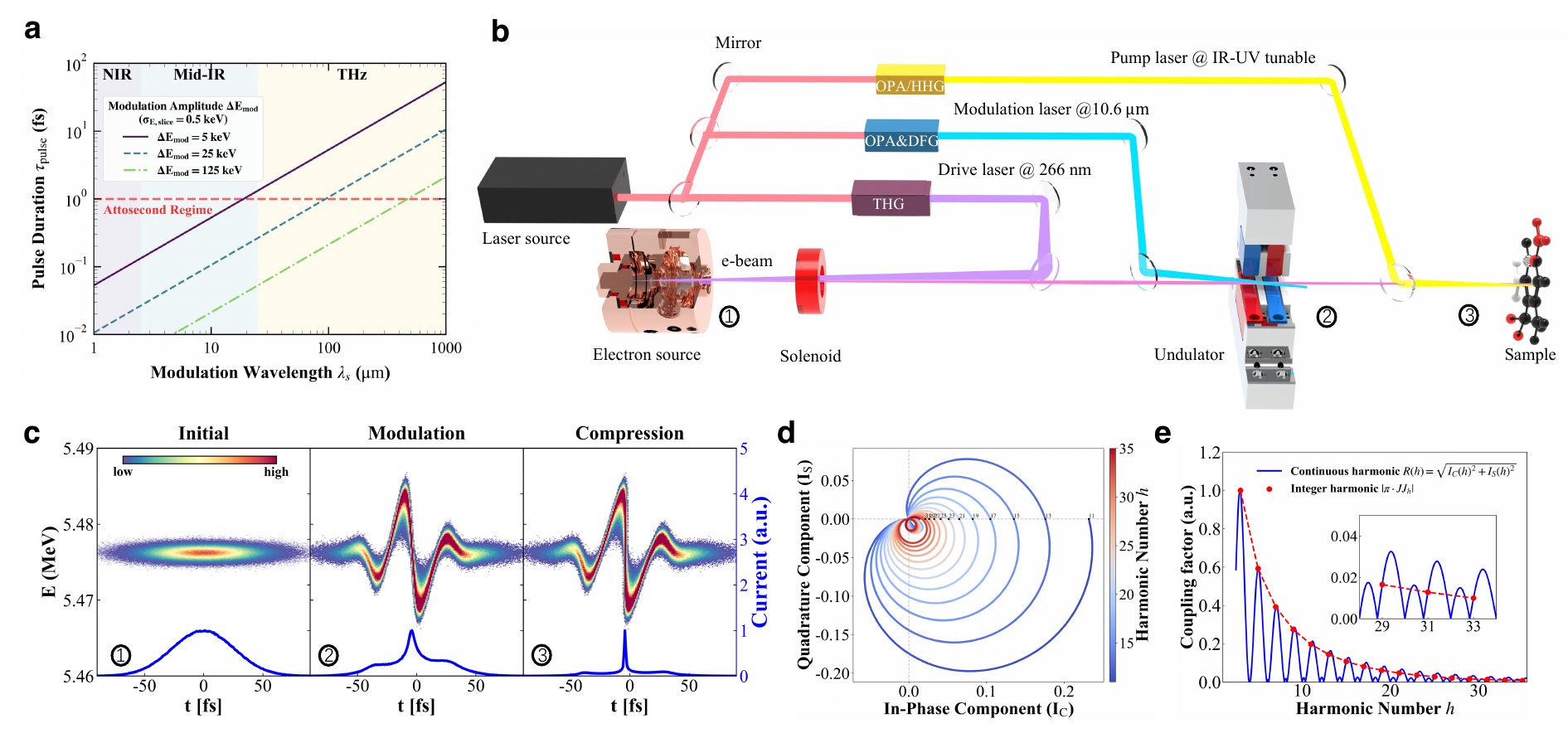}
\caption{
Conceptual layout and theoretical analysis of the all-optical manipulation for attosecond MeV electron pulse generation. (a) Linear scaling of the compressed electron pulse duration limit with modulation wavelength. At constant energy modulation amplitude $\Delta E_{\text{mod}}$ and slice energy spread $\sigma_{E,\text{slice}}$, the minimum pulse duration $\sigma_{t,\text{min}}$ scales linearly with the modulation wavelength $\lambda_s$. (b) Schematic of the all-optical attosecond ultrafast electron diffraction setup. Different colors in the diagram represent distinct laser branches derived from a single master laser: the pump laser for sample excitation (yellow), the laser for generating the probe electron beam (purple), and the mid-infrared laser for temporal modulation inside the undulator (blue). (c) Dynamic phase-space evolution of a representative MeV electron beam through the initial, modulation, and compression stages. (d) Phasor plot illustrating the trajectory of the in-phase ($I_C$) versus quadrature ($I_S$) components as the harmonic number $h$ varies continuously from 11 (outer spiral) to 31 (inner spiral). This spiral path explicitly demonstrates that the quadrature component $I_S$ is non-zero for non-integer $h$. (e) Generalized coupling factor as a function of the continuous harmonic number $h$ ($K = 2$). The blue solid line represents the full continuous solution $R(h)=\sqrt{I_C(h)^2+I_S(h)^2}$, which oscillates, while the red dashed line with dots represents the simplified integer-only theory acting as a smooth envelope connecting the odd-harmonic peaks.}
\label{fig1}
\end{figure*}

Various approaches have been developed to advance the temporal resolution of MeV electron beams. Radio-frequency (RF)–based velocity bunching has been demonstrated to compress MeV electron pulses to the 30 fs level \cite{PhysRevLett.118.154802,PhysRevX.8.021061}, with the potential to reach sub-femtosecond durations \cite{PhysRevAccelBeams.28.012802}. However, this approach is fundamentally limited by RF--laser synchronization jitter, which introduces arrival-time fluctuations between the compressed electron bunch and the pump laser, thereby constraining the achievable temporal resolution \cite{filippettoUltrafastElectronDiffraction2022}. Magnetic compression techniques, employing isochronous beamline conditions, have enabled simultaneous reduction of bunch length and timing jitter to the ~30 fs scale \cite{kimJitterfreeUltrafastElectron2020, qiBreaking50Femtosecond2020}. To further improve temporal resolution, laser-based compression schemes have been proposed, utilizing THz fields \cite{snivelyFemtosecondCompressionDynamics2020,PhysRevAccelBeams.21.082801,PhysRevLett.124.054802}, radially polarized lasers \cite{PhysRevApplied.17.064012}, or combined THz--optical modulation \cite{Lim_2019}. By deriving the modulation field from the same master oscillator as the pump pulse, these approaches establish intrinsic synchronization between the compressed electron beam and the excitation laser. As a result, THz-driven schemes have experimentally reduced both bunch duration and timing jitter to the ~30-fs level, while optical-wavelength modulation schemes theoretically pave the way for sub-femtosecond MeV electron pulses.

While THz-based methodologies have successfully pushed electron compression to the few-femtosecond frontier, breaking into the attosecond regime requires migrating to shorter modulation wavelengths—a strategy already validated in sub-femtosecond keV electron beams \cite{PhysRevLett.125.193202}. In free-space ballistic compression, the ultimate compressed bunch duration is fundamentally constrained by the initial energy spread and the modulation gradient. As dictated by phase-space dynamics (see Supplemental Material I \cite{SM}\nocite{kim2017synchrotron, saldin1999physics,stupakov2018classical}), the minimum compressed duration scales strictly linearly with the modulation wavelength ($\sigma_{t,min} \propto \lambda_L$) for a given slice energy spread and modulation amplitude [Fig.~\ref{fig1}(a)]. This linear scaling unlocks a critical advantage: by shifting to the mid-infrared (mid-IR) band, attosecond compression requires mere kiloelectronvolts of energy modulation. For MeV beams, this translates to a minimal relative energy spread (at $10^{-3}$ level), which effectively circumvents the severe chromatic aberrations that plague high-amplitude compression schemes. Advancing the driving field from the THz to the mid-IR band thus emerges as a natural and compelling pathway to surpass the conventional femtosecond limit.

In this Letter, we propose and numerically demonstrate an all-optical architecture for attosecond MeV-UED. Physically coupling mid-IR lasers ($\sim$$\upmu$m) to few-MeV electron beams inside a technologically mature macroscopic undulator ($\sim$cm) typically presents a massive wavelength mismatch, a fundamental barrier that we circumvent by generalizing the harmonic interaction to a continuous domain. To generate this 10.6 $\upmu$m modulation field while strictly eliminating the timing jitter inherent to RF environments, the entire setup is rooted in a single 800 nm Ti:sapphire master oscillator. As illustrated in Fig.~\ref{fig1}(b), parallel optical branches derived from this common source serve three distinct functions. The first branch drives the photocathode via third-harmonic generation ($266\text{ nm}$) to produce the high-brightness MeV electron bunch. The second branch feeds optical parametric amplification (OPA) and difference frequency generation (DFG) \cite{Petrov2015} stages to synthesize the crucial 10.6 $\upmu$m modulation laser. The final branch provides a versatile sample excitation pump, which can be derived from an OPA for wavelength tunability or from high-harmonic generation (HHG) to achieve attosecond temporal resolution. Because the 10.6 $\upmu$m modulation laser and the pump pulse originate from the same master oscillator, they remain intrinsically synchronized. Figure~\ref{fig1}(c) tracks the resulting longitudinal phase-space evolution. Co-propagating the electron bunch and the 10.6 $\upmu$m laser through a single-period undulator imprints a pure optical-cycle energy chirp. A subsequent free-space drift drives velocity bunching, converting this energy modulation into an isolated attosecond density modulation.

Conventional analyses of laser-beam interactions in an undulator typically restrict the harmonic resonance to the discrete integer regime ($h=n$) \cite{harmonic_PhysRevE.72.016501,RevModPhys.86.897}. To address this, we develop a universal theoretical framework by extending the interaction model to a continuous variable $h$, defined as the ratio of the fundamental resonant wavelength to the modulation laser wavelength ($h=\lambda_r/\lambda_s$). As detailed in Supplemental Material II \cite{SM}, the phase-dependent dimensionless energy modulation is expressed as:
\begin{equation}
M(\psi_0) = I_C \cos(\psi_0) + I_S \sin(\psi_0),
\end{equation}
where $\psi_0$ is the initial electron-laser phase, and the interaction is governed by two orthogonal coupling integrals:
\begin{subequations}
\begin{align}
I_C &= \int_0^{2\pi} \sin(\zeta)\sin(h \zeta-\alpha\sin 2\zeta) \,d\zeta, \\
I_S &= \int_0^{2\pi} \sin(\zeta)\cos(h \zeta-\alpha\sin 2\zeta) \,d\zeta.
\end{align}
\end{subequations}
Here, $\zeta=k_u z$ is the normalized longitudinal position within the undulator (where $k_u = 2\pi/\lambda_u$ is the undulator wavenumber and $z$ is the longitudinal coordinate), and $\alpha = h K^2 / (4(1+K^2/2))$, with $K$ representing the undulator parameter. The generalized coupling factor is thereby explicitly defined by the vector sum
\begin{equation}
R(h,K) = \sqrt{I_C^2 + I_S^2}.
\end{equation}
This generalization accounts for arbitrary detuning (non-integer $h$) and uncovers a crucial, previously overlooked quadrature component ($I_S$) arising from the broken symmetry of the interaction phase. Geometrically, the evolution of the phasor $(I_C, I_S)$ with varying $h$ is visualized in Fig.~\ref{fig1}(d). The resulting spiral morphology explicitly demonstrates that the quadrature component $I_S$ persists for all non-integer $h$, vanishing only at the discrete integer intercepts (black dots) where trigonometric orthogonality is satisfied over the $2\pi$ period. In this resonant limit ($I_S=0$), the interaction collapses to the purely in-phase regime, and Eq.~(2a) elegantly recovers the classical discrete Bessel factor for a planar undulator \cite{Harmonic_15th}:
\begin{equation}
I_C = \pi \left[ J_{\frac{h-1}{2}}(\alpha) - J_{\frac{h+1}{2}}(\alpha) \right] \equiv \pi JJ_h(\alpha).
\end{equation}
As depicted in Fig.~\ref{fig1}(e), the continuous coupling factor $R(h,K)$ forms a smooth envelope perfectly encompassing the discrete classical limits $|\pi JJ_h|$. At a macroscopic level, evaluating this envelope under a constant transverse driving parameter $K$ reveals a clear physical trend: coupling to progressively higher-order harmonics implies an intensified phase slippage across multiple optical cycles, which naturally drives the evident decay in the overall coupling efficiency. Crucially, however, at high-order harmonics, the maxima of this continuous envelope systematically deviate from exact integer resonances, demonstrating that a detuned resonance leverages the quadrature component to surpass conventional coupling limit.

Building upon this generalized framework, the absolute energy modulation amplitude $\Delta E_{mod}$ is defined as:
\begin{equation}
\Delta E_{mod} = \left( \frac{e E_s \lambda_u}{2\pi\gamma} \right) K \cdot R(h,K),
\end{equation}
where $e$ is the elementary charge, $E_s$ is the peak electric field of the modulation laser, $\lambda_u$ is the undulator period, and $\gamma$ is the relativistic Lorentz factor of the electron beam (see Supplemental Material II for detailed derivations). To elegantly balance the complex parameter space associated with this absolute modulation yield, a technologically mature $23\text{ mm}$ undulator is implemented for the subsequent optimization. This periodicity affords a sufficiently large $K$ to enhance the energy exchange, whilst intrinsically restricting the transverse trajectory ($x_{\mathrm{max}} = \frac{K \lambda_u}{2\pi\gamma}$) to maintain vital spatial coupling with the modulation laser core. Ultimately, this choice ensures a robust physical interaction while bypassing the prohibitive peak magnetic fields associated with extreme gap constraints. By fixing the undulator period, the beam energy, and assuming a constant peak laser field, the prefactor algebraically reduces to a purely constant. Consequently, the evolution of the modulation amplitude is strictly governed by the product of $K$ and the continuous coupling factor $R(h,K)$.

\begin{figure} [htbp]
\includegraphics[width=1\linewidth]{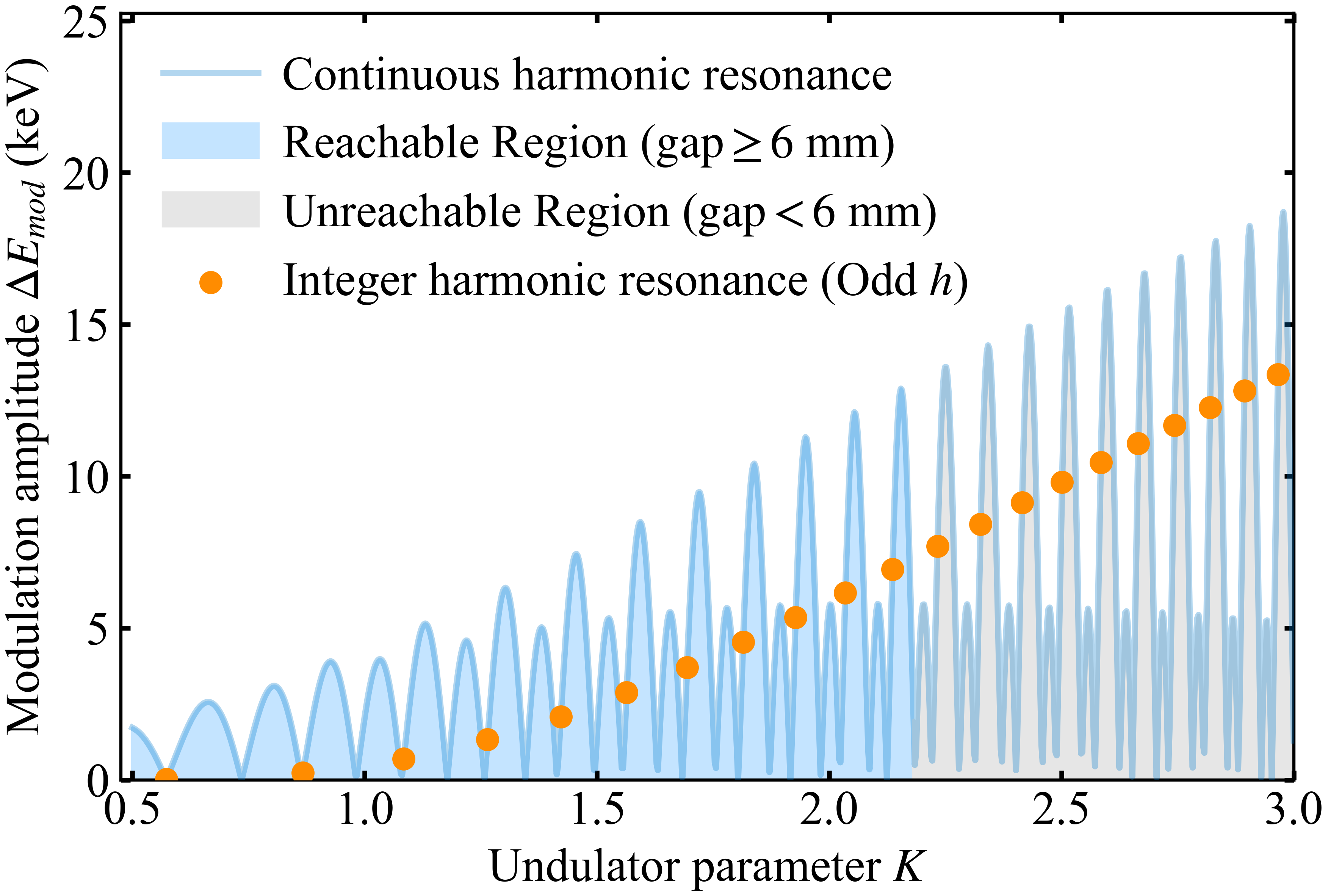}
\caption{Absolute modulation amplitude $\Delta E_{mod}$ as a function of the undulator parameter $K$ for $\lambda_u=23$ mm ($\gamma=10.716$), assuming a peak laser electric field of $E_s = 4.7 \times 10^8$ V/m. The blue solid curve represents the continuous-harmonic modulation theory, while the orange dots represent the discrete integer harmonic theory. The light blue shaded region marks the physically reachable parameter space with an undulator gap $\geq 6$ mm, and the gray region indicates the unreachable regime ($\mathrm{gap} < 6$ mm). The continuous theory forms a smooth envelope that exactly encapsulates all discrete harmonic solutions, demonstrating its unified and complete description of the resonant interaction.}
\label{fig2}
\end{figure}

Mapping this theoretical scaling onto the physical parameter space, Fig.~\ref{fig2} visualizes this amplitude as a function of $K$ for $\gamma = 10.716$. The solid light blue curve, rigorously derived from the continuous-harmonic theory, accurately captures the resonant electron--laser interaction over the full accessible parameter range. The orange discrete points, calculated from conventional integer-harmonic theory, lie entirely within the smooth envelope defined by the continuous-harmonic framework, highlighting its more complete and unified description of the harmonic interaction. Unlike the idealized constant-$K$ scenario, operating at a fixed undulator period dictates that $h$ and $K$ are strictly interlocked. For this specific configuration, the linearly increasing transverse kinematic velocity (represented by $K$) effectively dominates the interaction dynamics, driving a pronounced net upward trend in the overall modulation efficiency. Shaded regions demarcate the valid parameter space: the blue area corresponds to an undulator gap $\geq 6$ mm, while the gray region is inaccessible due to insufficient magnetic gap ($\mathrm{gap} < 6$ mm). Within the achievable $K$ range, the optimal continuous-harmonic operating point ($K=2.1542$, $h=31.37$) delivers an 85.3\% stronger modulation than the adjacent integer harmonic ($K=2.1361$, $h=31$). This significant performance gain over integer harmonic resonance is fundamentally driven by the constructive interference of the quadrature component $I_S$, which enhances the total interaction strength. Furthermore, by operating at a local maximum where the first derivative vanishes, unlike integer harmonic matching that constrains $K$ to the sensitive slope of the curve, our continuous framework endows the modulation efficiency with enhanced parameter robustness, thereby mitigating the impact of minor deviations in $K$.

\begin{figure} [htbp]
\includegraphics[width=1\linewidth]{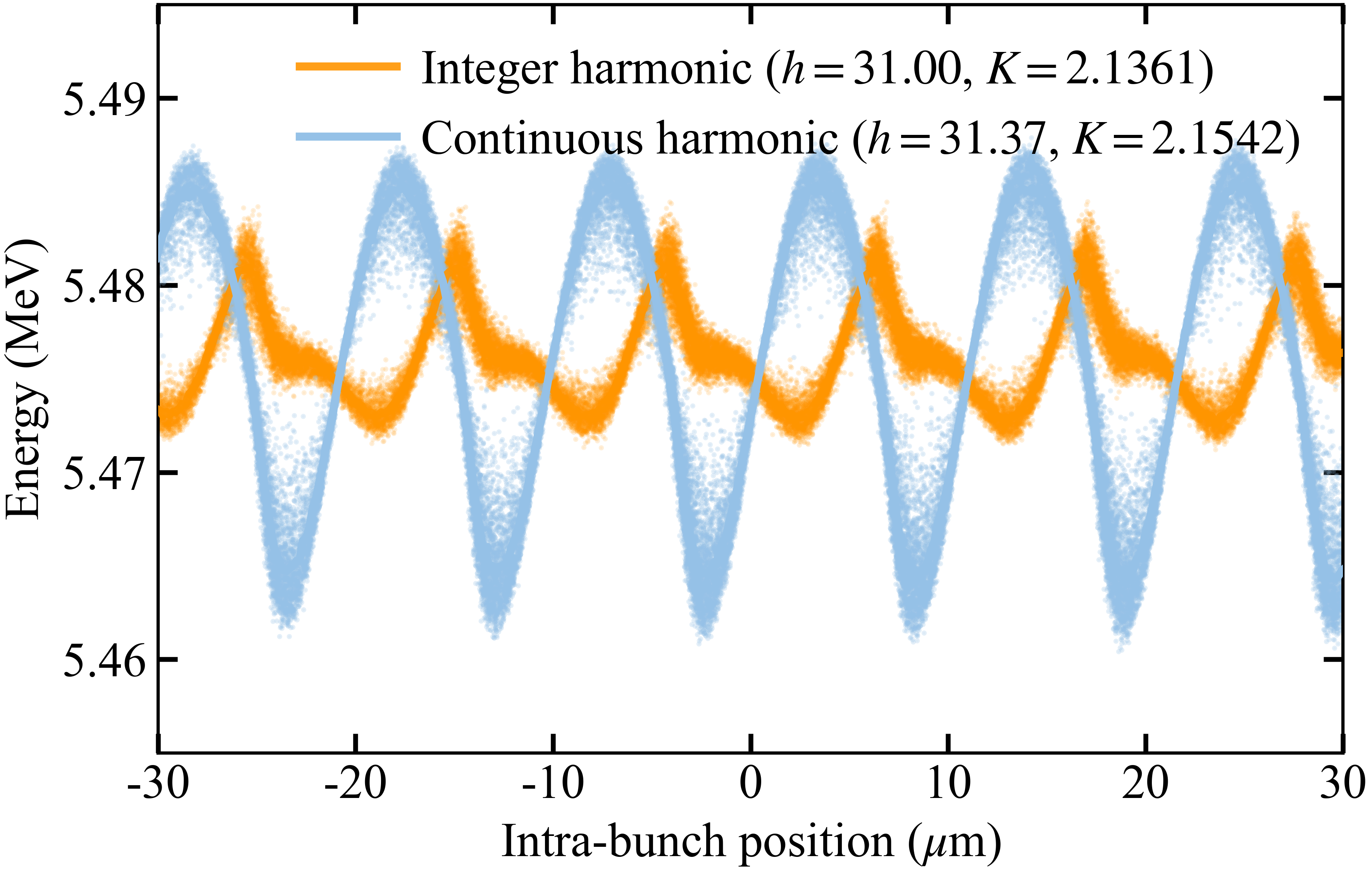}
\caption{Electron beam phase space under conventional integer-harmonic and optimized continuous-harmonic resonances. The orange curve represents the integer-harmonic operating point at $K=2.1361$ ($h=31$), and the lightblue curve denotes the optimized continuous-harmonic operating point at $K=2.1542$ ($h=31.37$). The continuous-harmonic scheme enhances the energy modulation depth by 83.7\% compared with the convential integer-harmonic method,
highlighting its remarkable advantage in resonance efficiency.}
\label{fig3}
\end{figure}

To quantitatively investigate laser-electron interactions in the continuous-harmonic regime, the dynamic performance is evaluated using three-dimensional particle-tracking simulations~\cite{PhysRevAccelBeams.22.070701}. The numerical model utilizes a $5.48\text{ MeV}$ electron beam characterized by a normalized emittance of $0.01\text{ mm}\cdot\text{mrad}$, with an intrinsic transverse size of $215\ \upmu\text{m}$ and a 0.01\% slice energy spread. The comprehensive 3D dynamics of the phase-space manipulation are fully resolved by simulating the interaction between an electron bunch (130 fs rms, 20 fC) and a focused mid-infrared modulation laser (wavelength: $10.6\ \upmu\text{m}$, pulse duration: 3 ps FWHM, peak power: 0.3 GW, laser waist size: 0.8 mm). To bridge the theoretical optimization with 3D numerical reality, a systematic parameter scan was conducted in the simulation by varying the undulator peak magnetic field at a fixed period of $\lambda_u = 23\ \text{mm}$ (one period). Figure~\ref{fig3} illustrates the central-region longitudinal phase-space distributions of the modulated electron beam for representative values of $K$. Crucially, the periodicity of this pronounced energy modulation is strictly dictated by the wavelength of the modulation laser, rather than the fundamental resonant wavelength. At the integer resonance ($K = 2.1361$, $h = 31$), corresponding to a magnetic gap of $6.111\text{ mm}$, the modulation amplitude is limited to $4.9\text{ keV}$. In contrast, the optimized continuous-harmonic configuration ($K=2.1542$, $h=31.37$) achieves an enhanced modulation amplitude of $9.0$ keV at a magnetic gap of $6.066$ mm. While the 1D theoretical model evaluates the interaction at the peak electric field $E_s$ to predict an upper limit of $\sim 12.8\text{ keV}$ (Fig.~\ref{fig2}), the complete 3D simulation rigorously accounts for the transverse spatial overlap between the finite-size electron beam and the Gaussian laser. Consequently, the attenuated modulation field experienced by off-axis electrons suppresses the macroscopically averaged modulation to the observed $9.0\text{ keV}$. Nevertheless, the numerically obtained relative modulation enhancement of $83.7\%$ is in excellent agreement with the theoretical scaling predictions, thereby validating the effectiveness of the proposed continuous harmonic optimization framework.

\begin{figure*} [htbp]
\includegraphics[width=18 cm]{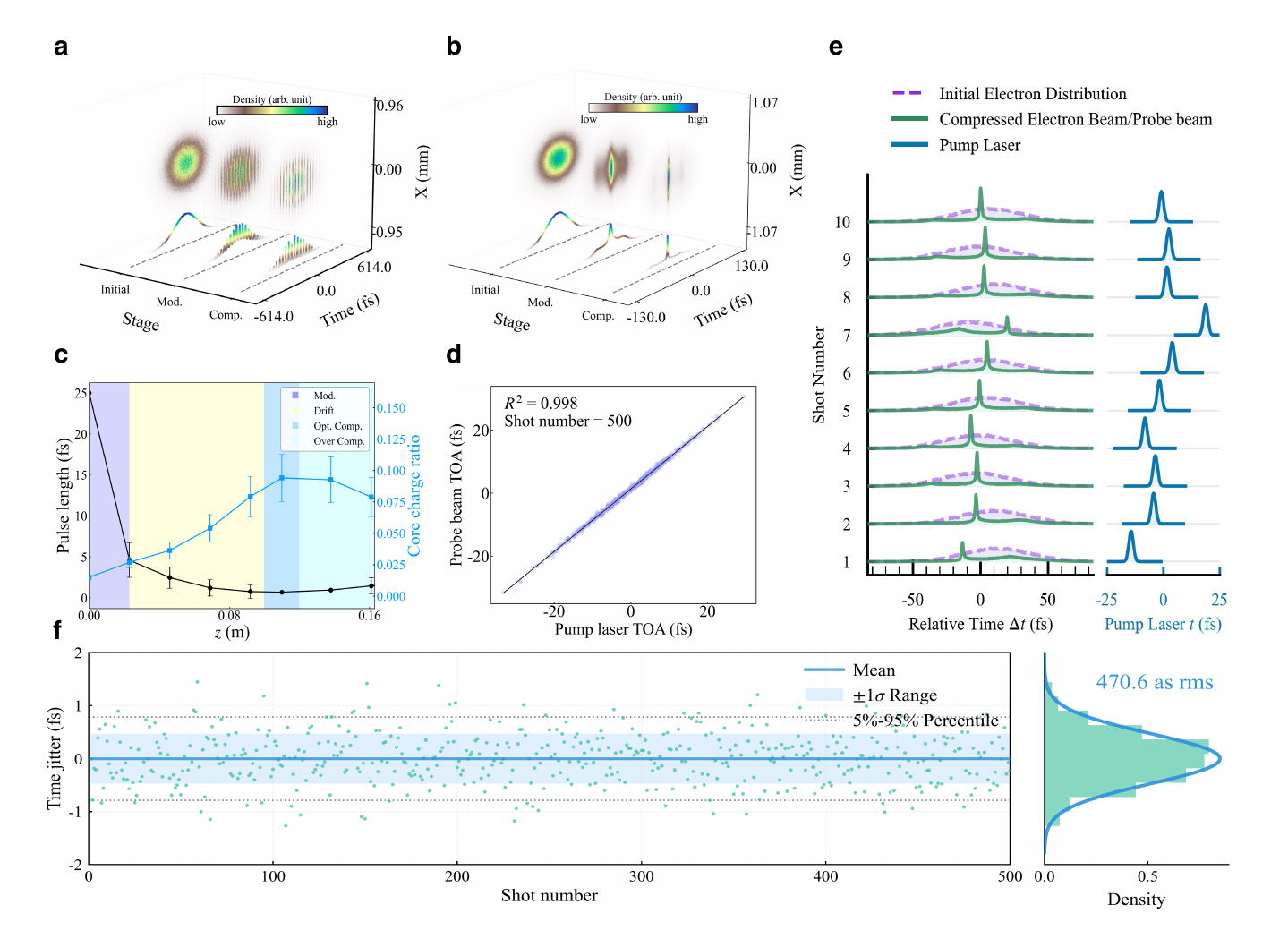}
\caption{Attosecond compression dynamics and intrinsic timing jitter suppression. (a) Spatiotemporal evolution demonstrating the generation of attosecond pulse trains from a longer initial electron bunch. (b) Spatiotemporal ($x-t$) evolution and corresponding current profiles of an isolated attosecond electron pulse at the initial, modulation, and compression stages. (c) Electron pulse duration (rms) and core charge ratio as a function of propagation distance. The shaded regions indicate the modulation stage (purple, Mod.), the drift section (yellow, Drift), the optimal compression regime (blue, Opt. Comp.), and the over-compression regime (cyan, Over Comp.). (d) Correlation between the time of arrival (TOA) of the compressed probe electron beam and the excitation pump laser. (e) Synchronous temporal locking of the pump laser and the compressed electron probe under realistic machine fluctuations, visualized using 10 representative shots. (f) Statistical distribution of the relative arrival-time jitter evaluated over a 500-shot ensemble, yielding an ultra-low relative jitter of 470.6 as (rms).
}
\label{fig4}
\end{figure*}

Harnessing the optimal modulation achieved in the continuous-harmonic regime, the electron beam acquires a quasi-linear energy chirp on the scale of the modulation laser wavelength. Driven by the enhanced modulation amplitude of $9.0 \text{ keV}$, subsequent propagation through an $8.7\text{ cm}$ free-space drift section initiates ballistic longitudinal velocity bunching. The corresponding evolution of the transverse density distributions---for the initial, modulated, and fully compressed electron beams---is illustrated in Fig.~\ref{fig4}(a). As this energy-time correlation rigidly rotates in the longitudinal phase space, the initial energy modulation is mapped into an ultra-dense spatial compression at the focal plane, yielding a well-defined attosecond pulse train with a duration of $680\text{ as}$ (rms).  Crucially, this highly efficient compression mechanism corroborates the core physical picture in Fig.~\ref{fig1}(a), demonstrating that even a modest energy modulation from a mid-infrared laser is sufficient to achieve attosecond compression while strictly preserving the beam quality required for high-fidelity diffraction.

In addition to producing attosecond pulse trains, the proposed scheme further enables the generation of isolated attosecond pulses. To suppress sideband generation and isolate a single density spike, a $35\text{ fs}$ (FWHM) modulation laser is utilized in conjunction with a shorter initial electron bunch (25 fs rms, 5 fC). The localized energy chirp induced by this ultra-short optical field drives the subsequent ballistic compression, yielding an isolated attosecond pulse of 688 as (rms). The transverse density distributions for the initial, modulated, and fully compressed electron beams in this isolated-pulse regime are presented in Fig.~\ref{fig4}(b), which closely echo the corresponding longitudinal phase-space evolution delineated in Fig.~\ref{fig1}(c). While the phase-space dynamics of attosecond compression are well established under ideal conditions, experimental viability hinges on the system's resilience to inherent machine fluctuations. To rigorously evaluate this, comprehensive 500-shot simulations are performed incorporating four primary fluctuation sources that dominantly dictate the longitudinal phase-space manipulation: initial electron energy jitter ($0.02\%$ rms), electron arrival-time jitter ($10$ fs rms), laser peak power jitter ($5\%$ rms), and laser arrival-time jitter ($10$ fs rms). The evolution of the pulse length and the core charge ratio (defined as the charge fraction within a 1 fs window) along the beamline distance $z$ is quantified in Fig.~\ref{fig4}(c). Optimal ballistic compression is achieved at a drift length of approximately 8.7 cm (blue shaded region), yielding a minimum bunch duration of $688.6 \pm 16.0$ as (rms) and a core charge ratio of $9.4\% \pm 1.9\%$. Such impressively small output fluctuations, particularly in the presence of 10-fs-level initial jitters, validate the exceptional robustness and experimental viability of the proposed scheme.

In the attosecond regime, the core charge is fundamentally constrained by severe longitudinal space-charge forces (LSC). As demonstrated by Maxson \textit{et al.} \cite{PhysRevLett.118.154802}, achieving sub-10 fs resolution in relativistic beams necessitates a strategic reduction in bunch charge (typically to the fC level) to circumvent LSC-induced energy spread growth and phase-space curvature. Rather than aiming for high single-shot charge—which inevitably degrades the temporal resolution—the proposed framework prioritizes the preservation of the sub-femtosecond temporal structure. By constraining the initial charge to $\sim$5 fC, the LSC-induced energy spread is confined to the tens-of-eV level (see Supplemental Material III \cite{SM}). Since this perturbation is two orders of magnitude smaller than the $\sim$9 keV laser induced modulation, its impact on the longitudinal phase-space linearity is rendered negligible, ensuring high-fidelity ballistic compression. Even as the charge scales to $20$ fC, the LSC-induced energy spread is held firmly beneath the $10\%$ modulation limit, providing a robust safety margin that shields the intricate attosecond structure from non-linear degradation. 

The performance of ultrafast electron probes in MeV-UED is fundamentally determined by the instrumental temporal resolution $\tau$ \cite{LI2009243}, which is governed by
\begin{equation}
\tau = \sqrt{\tau_{\text{pump}}^2 + \tau_{\text{probe}}^2 + \tau_{\text{jitter}}^2 + \tau_{\text{vm}}^2}.
\label{eq:resolution}
\end{equation}
With pump laser durations ($\tau_{\text{pump}}$) advancing into the sub-femtosecond regime---even reaching tens of attoseconds \cite{RevModPhys.81.163, doi:10.34133/ultrafastscience.0128}---and the velocity mismatch ($\tau_{\text{vm}}$) being minimally contributive for ultrathin samples owing to the highly relativistic electron velocity, the relative arrival-time jitter ($\tau_{\text{jitter}}$) between the pump and probe pulses emerges as the primary resolution bottleneck. In conventional photocathode RF guns, RF phase and amplitude fluctuations translate into severe timing jitter in the compressed beam \cite{PhysRevLett.118.154802}. The proposed all-optical manipulation scheme inherently circumvents this limitation. Driven by a homologous laser architecture---wherein both the modulation laser and the sample-excitation pump laser are optically split from a common master oscillator---the compressed attosecond electron pulses become intrinsically phase-locked to the pump pulses. This strict optical synchronization is evidenced by the near-perfect linear correlation ($R^2 = 0.998$) between their arrival times, as visualized in Fig.~\ref{fig4}(d). The shot-to-shot robustness of this mechanism under realistic jitter conditions is further elucidated in Fig.~\ref{fig4}(e). By extracting 10 typical shots from the 500-shot ensemble, this panel visualizes the synchronous locking of the compressed probe electron beam and the pump laser. Despite severe absolute timing fluctuations, their relative temporal separation remains rigidly locked. Consequently, the relative arrival-time jitter is drastically suppressed. Statistical analysis over the entire 500-shot ensemble, presented in Fig.~\ref{fig4}(f), yields an ultra-low relative jitter of $470.6\ \text{as}$ (rms). Such unprecedented temporal stability fundamentally enables the multi-shot diffraction signal accumulation necessitated by fC-level bunch charges, ensuring that the temporal coherence remains pristine and free from jitter-induced blurring throughout prolonged data acquisition. Integrating this minimal jitter with the achieved probe duration ($\tau_{\mathrm{probe}} \approx 688.6$ as rms) and a 200 as (rms) pump pulse achievable with current state-of-the-art technology \cite{RevModPhys.81.163, doi:10.34133/ultrafastscience.0128} establishes an overall system temporal resolution of $\tau \approx 858$ as (rms). This comprehensive evaluation validates the resilience of the all-optical scheme and demonstrates its capability to break the sub-femtosecond limit, paving the way for attosecond MeV-UED.

In conclusion, an all-optical phase-space manipulation approach based on continuous undulator harmonic resonance is proposed to simultaneously compress relativistic electron pulses into the attosecond regime and suppress timing jitter. Generalizing the undulator harmonic interaction to a continuous domain reveals a previously overlooked quadrature component ($I_S$) that drives constructive interference, enhancing the modulation amplitude upon detuning from the integer harmonic. Furthermore, operating at a local maximum where the first derivative vanishes provides intrinsic robustness against minor deviations in $K$, thereby relaxing stringent undulator field precision requirements and ensuring high resilience to shot-to-shot beam energy jitter. Operating at mid-infrared modulation wavelengths, the continuous harmonic approach overcomes the femtosecond compression barrier and generates attosecond electron pulses at modest modulation amplitudes ($10^{-3}$ level energy spread for MeV electron beams), strictly preserving the beam quality necessary for high-fidelity diffraction.  Simultaneously, the compressed electron probe is intrinsically phase-locked to the excitation pump lasers, suppressing the relative arrival time jitter to 470.6 as (rms). Accordingly, the overall temporal resolution is evaluated to reach the sub-femtosecond (rms) regime, assuming a state-of-the-art attosecond pump pulse. This sub-femtosecond temporal resolution extends the capabilities of MeV-UED, facilitating a transition from femtosecond-scale observations to the direct visualization of light-matter interactions at fundamental electronic timescales.

\textit{Acknowledgments}—This work is supported by the National Key R\&D Program of China (Grant No. 2024YFA1612200) funded by the Ministry of Science and Technology of the People's Republic of China, the National Natural Science Foundation of China (Grant No. 12505367, Grant No. 22288201), the Scientific Instrument Developing Project of Chinese Academy of Sciences (Grant No. GJJSTD20220001), the LiaoNing Revitalization Talents Program (Grant No. XLYC2202030) and the Strategic Priority Research Program of the Chinese Academy of Sciences (Grant No. XDB0970000, subject No. XDB0970100).

\textit{Data availability}—The data that support the findings of this article are openly available at~\cite{url2}.

\nocite{*}
\bibliography{references}
\end{document}